\documentclass[pra,aps,showpacs,twocolumn,twoside,superscriptaddress]{revtex4}

\usepackage{amsmath,amsfonts,amssymb,color,epsfig,graphics,graphicx,latexsym,revsymb,theorem,url,verbatim}

\usepackage{hyperref}

\newtheorem{definition}{Definition}
\newtheorem{proposition}[definition]{Proposition}
\newtheorem{lemma}[definition]{Lemma}

\newtheorem{theorem}[definition]{Theorem}
\newtheorem{corollary}[definition]{Corollary}
\newtheorem{conjecture}[definition]{Conjecture}

\newtheorem{remark}[definition]{Remark}
\newtheorem{example}[definition]{Example}

\def\squareforqed{\hbox{\rlap{$\sqcap$}$\sqcup$}}
\def\qed{\ifmmode\squareforqed\else{\unskip\nobreak\hfil
\penalty50\hskip1em\null\nobreak\hfil\squareforqed
\parfillskip=0pt\finalhyphendemerits=0\endgraf}\fi}
\def\endenv{\ifmmode\;\else{\unskip\nobreak\hfil
\penalty50\hskip1em\null\nobreak\hfil\;
\parfillskip=0pt\finalhyphendemerits=0\endgraf}\fi}
\newenvironment{proof}{\noindent \textbf{{Proof.~} }}{\qed}
\def\Dbar{\leavevmode\lower.6ex\hbox to 0pt
{\hskip-.23ex\accent"16\hss}D}
\makeatletter
\def\url@leostyle{%
  \@ifundefined{selectfont}{\def\UrlFont{\sf}}{\def\UrlFont{\small\ttfamily}}}
\makeatother
\def\bcj{\begin{conjecture}}
\def\ecj{\end{conjecture}}
\def\bcr{\begin{corollary}}
\def\ecr{\end{corollary}}
\def\bd{\begin{definition}}
\def\ed{\end{definition}}
\def\bea{\begin{eqnarray}}
\def\eea{\end{eqnarray}}
\def\bem{\begin{enumerate}}
\def\eem{\end{enumerate}}
\def\bex{\begin{example}}
\def\eex{\end{example}}
\def\bim{\begin{itemize}}
\def\eim{\end{itemize}}
\def\bl{\begin{lemma}}
\def\el{\end{lemma}}
\def\bpf{\begin{proof}}
\def\epf{\end{proof}}
\def\bpp{\begin{proposition}}
\def\epp{\end{proposition}}
\def\br{\begin{remark}}
\def\er{\end{remark}}
\def\bt{\begin{theorem}}
\def\et{\end{theorem}}

\newcommand{\nc}{\newcommand}

\def\a{\alpha}
\def\b{\beta}
\def\g{\gamma}
\def\d{\delta}
\def\e{\epsilon}

\def\z{\zeta}

\def\t{\theta}
\def\i{\iota}
\def\k{\kappa}
\def\l{\lambda}
\def\m{\mu}

\def\x{\xi}
\def\p{\pi}
\def\r{\rho}
\def\s{\sigma}

\def\ph{\varphi}

\def\ps{\psi}

\def\G{\Gamma}

\def\Ps{\Psi}

\nc{\bbC}{{\mathbb{C}}}

\nc{\cA}{{\cal A}} \nc{\cB}{{\cal B}} \nc{\cC}{{\cal C}}
\nc{\cD}{{\cal D}} \nc{\cE}{{\cal E}} \nc{\cF}{{\cal F}}
\nc{\cG}{{\cal G}} \nc{\cH}{{\cal H}} \nc{\cI}{{\cal I}}
\nc{\cJ}{{\cal J}} \nc{\cK}{{\cal K}} \nc{\cL}{{\cal L}}
\nc{\cM}{{\cal M}} \nc{\cN}{{\cal N}} \nc{\cO}{{\cal O}}
\nc{\cP}{{\cal P}} \nc{\cR}{{\cal R}} \nc{\cS}{{\cal S}}
\nc{\cT}{{\cal T}} \nc{\cU}{{\cal U}} \nc{\cV}{{\cal V}}
\nc{\cW}{{\cal W}} \nc{\cX}{{\cal X}} \nc{\cZ}{{\cal Z}}

\nc{\hA}{{\hat{A}}} \nc{\hB}{{\hat{B}}} \nc{\hC}{{\hat{C}}}
\nc{\hD}{{\hat{D}}} \nc{\hE}{{\hat{E}}} \nc{\hF}{{\hat{F}}}
\nc{\hG}{{\hat{G}}} \nc{\hH}{{\hat{H}}} \nc{\hI}{{\hat{I}}}
\nc{\hJ}{{\hat{J}}} \nc{\hK}{{\hat{K}}} \nc{\hL}{{\hat{L}}}
\nc{\hM}{{\hat{M}}} \nc{\hN}{{\hat{N}}} \nc{\hO}{{\hat{O}}}
\nc{\hP}{{\hat{P}}} \nc{\hR}{{\hat{R}}} \nc{\hS}{{\hat{S}}}
\nc{\hT}{{\hat{T}}} \nc{\hU}{{\hat{U}}} \nc{\hV}{{\hat{V}}}
\nc{\hW}{{\hat{W}}} \nc{\hX}{{\hat{X}}} \nc{\hZ}{{\hat{Z}}}

\def\dim{\mathop{\rm Dim}}

\def\max{\mathop{\rm max}}
\def\min{\mathop{\rm min}}

\def\rank{\mathop{\rm rank}}

\def\sr{\mathop{\rm sr}}

\def\tr{\mathop{\rm Tr}}

\def\GL{{\mbox{\rm GL}}}

\def\dg{\dagger}

\def\ox{\otimes}

\newcommand{\bra}[1]{\langle#1|}
\newcommand{\ket}[1]{|#1\rangle}
\newcommand{\proj}[1]{| #1\rangle\!\langle #1 |}
\newcommand{\ketbra}[2]{|#1\rangle\!\langle#2|}
\newcommand{\braket}[2]{\langle#1|#2\rangle}
\newcommand{\norm}[1]{\lVert#1\rVert}
\newcommand{\abs}[1]{|#1|}

\newcommand{\jmp}{J. Math. Phys.}
\newcommand{\jpa}{J. Phys. A}

\def\bR{{\mbox{\bf R}}}

\def\bC{{\mbox{\bf C}}}

\begin{document}
\title{Non-positive-partial-transpose quantum states of rank four are distillable}

\author{Lin Chen}
\affiliation{School of Mathematics and Systems Science, Beihang University, Beijing 100191, China}
\affiliation{International Research Institute for Multidisciplinary Science, Beihang University, Beijing 100191, China} \email{linchen@buaa.edu.cn (corresponding author)}

\def\Dbar{\leavevmode\lower.6ex\hbox to 0pt
{\hskip-.23ex\accent"16\hss}D}
\author {{ Dragomir {\v{Z} \Dbar}okovi{\'c}}}

\affiliation{Department of Pure Mathematics and Institute for
Quantum Computing, University of Waterloo, Waterloo, Ontario, N2L
3G1, Canada} \email{djokovic@uwaterloo.ca}

\begin{abstract}
We show that any bipartite quantum state of rank four is distillable, when the partial transpose of the state has at least 
one negative eigenvalue, i.e., the state is NPT. For this purpose we prove that if the partial transpose of a two-qutrit NPT state 
has two non-positive eigenvalues, then the state is distillable. We further construct a parametrized two-qutrit NPT entangled state of rank five which is not 1-distillable, and show that it is not $n$-distillable for any given $n$ when the parameter is sufficiently small. This state has the smallest rank among all  1-undistillable NPT states. We conjecture that the state is not distillable.
\end{abstract}

\date{ \today }




\maketitle


\section{Introduction}

Many quantum-information tasks require entangled pure states as the necessary resource. In entanglement theory, a long-standing open problem is whether bipartite NPT states can be asymptotically converted into pure entangled states under local operations and classical communications (LOCC). This is the well-known distillability problem \cite{dss00} and is related to the super-activation of bound entanglement and zero capacity quantum channels. 

In spite of much effort devoted in past years, the distillability problem is still far from a solution due to the rapidly increasing number of parameters in the density matrices of states. A method of avoiding this difficulty is to convert by LOCC the states into the Werner states containing only one parameter 
\cite[section II]{dss00}. Thus to solve the distillability problem, it suffices to distill Werner states. In recent years there were several attempts to do that  \cite{dss00,Vianna,Pankowski,Hiroshima,Bandyopadhyay,Kraus,Clarisse}. 
On the other hand, progress towards distilling entangled states of given dimensions or deficient rank has been made steadily. Compared with Werner states, in the case of deficient rank one can make use of more fruitful features of of density matrices, such as the existence of product vectors in the kernel and range. Entangled states of rank two and three \cite{hst99,cc08}, $2\times N$ NPT states \cite{Horodecki2}, and $M\times N$ entangled states of rank at most $\max(M,N)$ \cite{hst99,cd11JPA} have been proven distillable. 

In this paper we show that if the partial transpose of an NPT 
two-qutrit state has at least two non-positive eigenvalues 
counting multiplicities, then the state is distillable. This is presented in Theorem \ref{thm:3x3distillable,rankrhoG<9}.
We further show that any NPT state of rank four is distillable. This is a corollary of the fact proved in Theorem \ref{thm:3x3distillable,prod=kernel}, that if the kernel of a two-qutrit NPT state contains a product vector, then the state is 
1-distillable.  
Since entangled states of rank at most three are distillable \cite{hst99,cc08}, any NPT state of rank at most four is distillable. So the distillable entanglement measure is positive for these states \cite{bds96}. Based on these facts, we give in 
Corollary \ref{cr:1nondistillable} some necessary conditions for a bipartite state not to be 1-distillable. We construct a 
two-qutrit NPT state $\r$ of rank five, see Eq. \eqref{eq:r=s}, 
depending on a parameter $\epsilon>0$, which is not 
1-distillable. This is achieved by using the positive-partial-transpose (PPT) entangled edge states constructed in \cite[p11]{ko12}, namely \eqref{eq:sigma} in this paper. Since any NPT state of rank at most four is 1-distillable, $\r$ has the smallest rank among all 1-undistillable NPT states. We further show in Lemma \ref{le:undistillable} that for any given integer $n$ and sufficiently small $\epsilon>0$, $\r$ is 
$n$-undistillable. This is based on some estimates for many-copy of separable Werner states, presented in Lemma \ref{le:NcopyWernerSEP}. We conjecture that $\r$ is not distillable 
for small $\epsilon>0$.

In the literature many results about two-qutrit states of rank four have been found by using the PPT states constructed from the unextendible product bases (UPB) \cite{bdm99}. Next, any such states were proved to be constructed by UPB up to stochastic LOCC (SLOCC) \cite{cd11JMP,dvc00}. Further, the sufficient and necessary condition of deciding the separable states of rank four has been proposed \cite{werner89,cd11JPA}. These facts and our result together classify all two-qutrit entangled states of rank four.

The rest of the paper is organized as follows. In Sec. \ref{sec:pre} we introduce the mathematical formulation of distillability problem and notation used in the paper. In Sec. \ref{sec:main} we present the conditions for 1-distillable and 1-undistillable two-qutrit NPT states, and in particular we show 
that all two-qutrit NPT states of rank four are 1-distillable. We also construct a 1-undistillable two-qutrit NPT states of rank five. In Sec. \ref{sec:Ndistillability} we construct 
n-undistillable two-qutrit NPT states of rank five for any integer $n$. We give our conclusions in Sec. \ref{sec:con}.

\section{preliminaries}
\label{sec:pre}

In this section we introduce the notation and recall facts on 
the distillability problem which will be used throughout 
the paper.
Let $\cH=\cH_A\ox\cH_B$ be the bipartite Hilbert space  with $\dim\cH_A=M$ and  $\dim\cH_A=N$. We shall work with bipartite quantum states $\r$ on $\cH$. We shall write $I_k$ for the identity $k\times k$ matrix. We denote by $\cR(\r)$ and $\ker \r$ the range and kernel of a linear map
$\r$, respectively. From now on, unless stated otherwise, the states will not
be normalized. We shall denote by $\{\ket{i}_A:i=0,\ldots,M-1\}$ and
$\{\ket{j}_B:j=0,\ldots,N-1\}$ o. n. bases of $\cH_A$ and $\cH_B$,
respectively. The partial transpose of $\r$ w. r. t. the system $A$ is defined as $\rho^{\G}:=\sum_{i,j}\ketbra{j}{i}\ox\bra{i}\r\ket{j}$. We say that $\r$ is PPT if $\r^{\G}\ge0$. Otherwise $\r$ is NPT, i.e., $\r^{\G}$ has at least one negative eigenvalue. The NPT states are always entangled due to the
Peres-Horodecki criterion for separable states \cite{Peres}.

The distillability problem requires many-copy states, so we introduce the notion of composite system.
Let $\r_{A_iB_i}$ be an $M_i\times N_i$ state of rank $r_i$ acting
on the Hilbert space $\cH_{A_i}\ox\cH_{B_i}$, $i=1,2$. Suppose
$\r$ of systems $A_1,A_2$ and $B_1,B_2$ is a state acting on the
Hilbert space
$\cH_{A_1}\ox\cH_{B_1}\ox\cH_{A_2}\ox\cH_{B_2}$, such that $\tr_{A_1B_1}\r=\r_{A_2B_2}$ and
$\tr_{A_2B_2}\r=\r_{A_1B_1}$. By
switching the two middle factors, we can consider $\r$ as a
\textit{composite} bipartite state acting on the Hilbert space
$\cH_A\ox\cH_B$ where $\cH_A=\cH_{A_1}\ox\cH_{A_2}$ and
$\cH_B=\cH_{B_1}\ox\cH_{B_2}$. In that case we shall write
$\r=\r_{A_1A_2:B_1B_2}$. So $\r$ is an $M_1M_2\times N_1N_2$
state of rank not larger than $r_1r_2$. In particular for the
\textit{tensor product} $\r=\r_{A_1B_1}\ox\r_{A_2B_2}$, it is easy
to see that $\r$ is an $M_1M_2\times N_1N_2$ state of rank $r_1r_2$.

The above definition can be easily generalized to the tensor product
of $N$ states $\r_{A_iB_i},i=1,\ldots,N$. They form a bipartite
state on the Hilbert space
$\cH_{A_1,\cdots,A_N}\ox\cH_{B_1,\cdots,B_N}$. It is written as $\cH^{\ox n}$ when $\cH_{A_i}\ox\cH_{B_i}=\cH$. Using this terminology, we introduce the definition of distillable states \cite{dss00}.
\bd
 \label{def:distillation}
A bipartite state $\r$ is {\em $n$-distillable} under LOCC if there
exists a Schmidt-rank-two state $\ket{\ps}\in\cH^{\ox n}$ such
that $\bra{\ps} ({\r^{\ox n}})^\G \ket{\ps}<0$. Otherwise we say
that $\r$ is $n$-undistillable. We say that $\r$ is {\em
distillable} if it is $n$-distillable for some $n\ge1$. If an entangled state $\r$ is not distillable, then we say that it is {\em bound entangled}.
 \ed

It is immediate from this definition that no PPT state is  distillable. Hence PPT entangled states are bound entangled states. The distillability problem asks whether bound entangled state can be NPT.

Let us recall some basic methods for
proving the separability, distillability and PPT properties of 
bipartite states. We say that two bipartite states $\r$ and $\s$ are equivalent
under SLOCC if there exists an invertible local operator
(ILO) $A\ox B$ such that $\r=(A^\dg \ox B^\dg) \s (A \ox B)$ \cite{dvc00}.
It is easy to see
that any ILO transforms distillable, PPT, entangled, or separable state into the
same kind of states. We shall often use ILOs to simplify the density
matrices of states. A subspace which contains no product state, 
is referred to as a completely entangled subspace (CES).

\section{1-distillability of NPT states}
\label{sec:main}

For convenience, we denote by $\sr(x)$ the Schmidt rank of a state
$\ket{x}\in\cH$.
If $\ket{x}=\sum_{i,j} \x_{ij}\ket{i,j}$, $0\le i<M$, $0\le j<N$,
we say that $[\x_{ij}]$ is the matrix of $\ket{x}$.
By definition, $\sr(x)$ is the ordinary rank of $[\x_{ij}]$. The following lemma gives a necessary condition for the 1-distillability of quantum states.

 \bl
 \label{le:PTeigenvector=entangled}
If $\r$ is a bipartite state and $\bra{\psi}\r^\G\ket{\psi}<0$
for some vector $\ket{\psi}$, then $\ket{\psi}$ is entangled.
 \el
 \bpf
Assume that $\ket{\ps}=\ket{f,g}$. Then
$\bra{\ps}\r^\G\ket{\ps}=\bra{f^*,g}\r\ket{f^*,g}\ge0$ gives a
contradiction.
 \epf

The following result generalizes \cite[Lemma 4]{cd11JPA}, and their proofs are also similar.
\bl
\label{le:2x2}
Let $\r$ be a bipartite state such that $\r^\G$ has a principal $2\times2$ submatrix of negative determinant. Then $\r$ is distillable.
\el
\bpf
Let $\r=\sum_{i,j}\ketbra{i}{j}\ox\s_{ij}$, and the $2\times2$ submatrix $\left[
           \begin{array}{ccc}
             a & b\\
             b^* & c
           \end{array}
         \right]$. We have $\r^\G=\sum_{i,j}\ketbra{i}{j}\ox\s_{ji}$. Since $\s_{ii}\ge0$, the diagonal entries a and c must belong to different diagonal blocks, say $\s_{kk}$ and $\s_{ll}$. Let P be the orthogonal projector onto the two-dimensional subspace of $\cH_A$ spanned by $\ket{k}$ and $\ket{l}$. Then the projected state $(P\ox I_B)\r(P\ox I_B)$ is an NPT state on the $2\times N$ system. Hence it is distillable, and $\r$ is also distillable.
\epf

We can now prove the main result of this section.
 \bt
 \label{thm:3x3distillable,rankrhoG<9}
If $\r$ is a two-qutrit NPT state and $\r^\G$ has at least two
non-positive eigenvalues counting multiplicities, then $\r$ is
1-distillable.
 \et
 \bpf
By the hypothesis, there exist two eigenvectors of $\r^\G$, 
say $\ket{\a}$ and $\ket{\b}$ with matrices $A$ and $B$, such 
that $\r^\G\ket{\a}=\lambda\ket{\a}$, $\lambda<0$, 
$\r^\G\ket{\b}=\mu\ket{\b}$, $\mu\le0$, and $\braket{\a}{\b}=0$. 
If $A$ is not invertible, then its rank is 2 and so $\r$ is
1-distillable.
From now on we assume that $A$ is invertible, and also that 
$B$ is invertible if $\mu<0$.
Assume that $N:=A^{-1}B$ is not nilpotent. Then $\det(I_3+tN)$ is a nonconstant polynomial in $t$ 
and we can choose $t$ so that this determinant is 0. 
Thus $A+tB$ is singular, and 
$\ket{\phi}:=\ket{\a}+t\ket{\b}$
satisfies 
$\bra{\phi} \r^\G \ket{\phi} = \bra{\a} \r^\G \ket{\a}
+|t|^2  \bra{\b} \r^\G \ket{\b}<0$. 
Hence $\r$ is 1-distillable.
If $N$ is nilpotent we can choose $\ket{\a'}$, with matrix $A'$,  close to $\ket{\a}$ such that $\bra{\a'}\r^\G\ket{\a'}<0$, and 
such that the matrix $N':=(A')^{-1}B$ is not nilpotent. 
Then we can use the previous argument to show that $\r$ is 
1-distillable. 
\epf

Using this result, we show that the existence of product vectors in the kernel is related to 1-distillability.

 \bt
 \label{thm:3x3distillable,prod=kernel}
If the kernel of a two-qutrit NPT state $\r$ contains a product
vector, then $\r$ is 1-distillable. Hence any bipartite NPT 
state $\r$ of rank four is 1-distillable.
 \et
 \bpf
We can assume that $\ket{0,0}\in\ker\r$. 
Consequently, the first diagonal entry of $\r$ is 0, 
and the same is true for $\r^\G$. 
If the first column of $\r^\G$ is not 0, then $\r$ is 
1-distillable by \cite[Lemma 4]{cd11JPA}. 
Otherwise  $\ket{0,0}\in\ker\r^\G$ and $\r$ is 1-distillable
by Theorem \ref{thm:3x3distillable,rankrhoG<9}. 
We have proved the first assertion.

To prove the second assertion, we recall that $M\times N$ NPT states of rank four are 1-distillable when $\max(M,N)\ge4$ \cite{hst99,cd11JPA} or $\min(M,N)=2$. It remains to consider the 
case $M=N=3$. Since $\dim\ker\r=5$, $\ker\r$ contains a 
product vector. Hence, the first assertion implies the second 
in this case.
\epf

Our results show that if we can convert an entangled state into an NPT state of rank four, then it is distillable. It provides a new way of attacking the distillability problem. 
Let us mention that the result proved in \cite{cc08} is the special
case of this theorem when $r=3$, and that \cite[Theorem 28]{cd11JPA}
follows from the case $r=4$.
It is immediate from the definition of bad states in \cite{cd13} that the kernel
of such a state contains at least one product vector.
Consequently, all bad two-qutrit NPT states are 1-distillable.
We may further construct 1-distillable two-qutrit NPT states $\r$ of
rank $r$ ranging from five to nine. Suppose $\s$ is a two-qutrit NPT
states of rank four. We can find five product vectors
$\ket{a_i,b_i},i=1,2,3,4,5$, such that they and $\cR(\s)$ span the
space $\bbC^3 \ox \bbC^3$. Then
$\r=\s+\e\sum^{r-4}_{i=1}\proj{a_i,b_i}$ is NPT and
1-distillable when $\e>0$ is sufficiently small.
In spite of these results, there exist 1-undistillable two-qutrit
NPT states of rank five. They satisfy the following statement in view of the above theorems.

 \bcr
 \label{cr:1nondistillable}
If $\r$ is a 1-undistillable two-qutrit NPT state, then $\ker\r$ is a CES, and $\r^\G$ has exactly one negative and eight
positive eigenvalues. Consequently, $\rank\r>4$ and $\det\r^\G\ne0$.
 \ecr

We analytically construct such $\r$.  We use the fact \cite[p10]{ko12} that there exists a 2-parameter family of two-qutrit
edge entangled states $\s^\G$ of birank $(8,5)=(\rank\s^\G,\rank\s)$ where
\begin{widetext}
\bea
\label{eq:sigma}
\s:={1\over 3(2\cos\t + b +1/b) }\left[
           \begin{array}{ccccccccc}
            2\cos\t & 0 & 0 & 0 & -\cos\t & 0 & 0 & 0 & -\cos\t \\
            0 & {1\over b} & 0 & -e^{-i\t} & 0 & 0 & 0 & 0 & 0 \\
            0 & 0 & b & 0 & 0 & 0 & -e^{i\t} & 0 & 0 \\
            0 & -e^{i\t} & 0 & b & 0 & 0 & 0 & 0 & 0 \\
            -\cos\t & 0 & 0 & 0 & 2\cos\t & 0 & 0 & 0 & -\cos\t \\
            0 & 0 & 0 & 0 & 0 & {1\over b} & 0 & -e^{-i\t} & 0 \\
            0 & 0 & -e^{-i\t} & 0 & 0 & 0 & {1\over b} & 0 & 0 \\
            0 & 0 & 0 & 0 & 0 & -e^{i\t} & 0 & b & 0 \\
            -\cos\t & 0 & 0 & 0 & -\cos\t & 0 & 0 & 0 & 2\cos\t \\
           \end{array}
         \right],
\eea
and thus
\bea
\label{eq:sigmaG}
\s^\G={1\over 3(2\cos\t + b +1/b) }\left[
           \begin{array}{ccccccccc}
            2\cos\t & 0 & 0 & 0 & -e^{i\t} & 0 & 0 & 0 & -e^{-i\t} \\
            0 & {1\over b} & 0 & -\cos\t & 0 & 0 & 0 & 0 & 0 \\
            0 & 0 & b & 0 & 0 & 0 & -\cos\t & 0 & 0 \\
            0 & -\cos\t & 0 & b & 0 & 0 & 0 & 0 & 0 \\
            -e^{-i\t} & 0 & 0 & 0 & 2\cos\t & 0 & 0 & 0 & -e^{i\t} \\
            0 & 0 & 0 & 0 & 0 & {1\over b} & 0 & -\cos\t & 0 \\
            0 & 0 & -\cos\t & 0 & 0 & 0 & {1\over b} & 0 & 0 \\
            0 & 0 & 0 & 0 & 0 & -\cos\t & 0 & b & 0 \\
            -e^{i\t} & 0 & 0 & 0 & -e^{-i\t} & 0 & 0 & 0 & 2\cos\t \\
           \end{array}
         \right],
\eea
\end{widetext}
where the two parameters $b>0$ and $0<\abs{\t}<\p/3$.
We have the spectral decomposition
\bea
\label{eq:sg}
\s^\G=\sum^8_{i=1}p_i\proj{\ph_i},
\eea
where $p_i$ are the
positive eigenvalues of $\s^\G$. Let $p_1=\min_i p_i$. By computation we have
\bea
p_1=&&
{1\over 6\cos\t + 3b +3/b }
\min \bigg\{ 3\cos\t-\sqrt3 \abs{\sin\t}, 
\notag\\
&&{1+b^2-\sqrt{1+b^4+2b^2\cos 2\t} \over 2b } \bigg\}.
\eea
Eq. \eqref{eq:sigma} implies that
$\ker\s^\G$ is spanned by the two-qutrit maximally entangled 
state 
\bea
\label{eq:mes}
\ket{\Ps}:={1\over\sqrt{3}}(\ket{00}+\ket{11}+\ket{22}).
\eea
Since $\rank\s=5$, $\cR(\s)$ contains a normalized product state $\ket{f,g}$. One can verify that we may choose
\bea
\ket{f,g}={1\over  b^{\frac12}  + b^{-\frac12}  }
(\ket{0}+b^{\frac12} e^{{i\over 2} \t} \ket{1})(\ket{0}-b^{-\frac12} e^{-{i\over 2} \t} \ket{1}),
\notag\\
\eea
such that $\ket{f^*,g}\not\in\cR(\r^\G)$.
For sufficiently small $\e\in(0,p_1/3]$, the matrix
\bea
\label{eq:r=s}
\r=\s-\e\proj{f,g}
\eea 
is positive semidefinite and satisfies $\rank\r=5$. So $\r$ is a non-normalized quantum state.
Since $\ket{f^*,g}\notin\cR(\s^\G)$, the partial transpose
$\r^\G=\s^\G-\e\proj{f^*,g}$ is not positive semidefinite. 
So $\r$ is NPT.

Let $\ket{\ps}$ be a normalized pure state of Schmidt rank 
at most two. Since $\s^\G\ge0$ and its kernel is spanned
by $\ket{\Ps}$, we have $\bra{\ps}\s^\G\ket{\ps}>0$. Since
\cite[Lemma 3]{dss00} implies
\bea
\label{eq:psps}
\max_{\ps}\abs{\braket{\Ps}{\ps}}^2=\frac23,
\eea
Eq. \eqref{eq:sg} implies that
 \bea
 \label{ea:lowerbound}
 \bra{\ps}\s^\G\ket{\ps}
 &=& \sum_i p_i \braket{\ps}{\ph_i}\braket{\ph_i}{\ps}
 \notag\\
 &\ge& p_1 \sum_i \braket{\ps}{\ph_i}\braket{\ph_i}{\ps}
 \notag\\
 &=& p_1 \bra{\ps} (I_9-\proj{\Ps}) \ket{\ps} 
 \notag\\
 &\ge& \frac{p_1}{3}.
 \eea
As $|\braket{f^*,g}{\ps}|<1$, we have
$\bra{\ps}\r^\G\ket{\ps}
=\bra{\ps}(\s^\G-\e\proj{f^*,g})\ket{\ps}>p_1/3-\e\ge0$.
Hence $\r$ is 1-undistillable.

\section{$N$-distillability of NPT states}
\label{sec:Ndistillability}

In this section we study the n-distillability of NPT states. We will construct a parametrized family of two-qutrit NPT states of rank five, which is not $n$-distillable when the parameter is a function of $n$ and small enough. This is similar to the phenomenon in \cite{dss00}. For this purpose we present a preliminary lemma. We conjecture that the lhs of \eqref{eq:NcopyWernerSEPmin} is equal to ${1\over2}\cdot{1\over 12^n}$.

 \bl
\label{le:NcopyWernerSEP}
Let $\r_s={1\over8}(I_9-\proj{\Ps})$. For any integer $n$ and any pure
state $\ket{\ps}$ of Schmidt rank at most two,
we have 
\bea
\label{eq:NcopyWernerSEP1}
&&
(\s^\G)^{\ox n}
\ge
{(8p_1)^n} (\r_s)^{\ox n},
\\
&&
\label{eq:NcopyWernerSEPmax}
\max_{\ps}\bra{\ps}(\r_s)^{\ox n}\ket{\ps}={1 \over 8^n},
\\
&&
\label{eq:NcopyWernerSEPmin}
\min_{\ps}\bra{\ps}(\r_s)^{\ox n}\ket{\ps} \ge {1 \over 24^n}.
\eea
 \el
\bpf
Eq. \eqref{eq:sg} implies
\eqref{eq:NcopyWernerSEP1}. Next, \eqref{eq:NcopyWernerSEPmax} follows from the fact that $\r_s^{\ox n}$ is a normalized projector of rank $8^n$, and the maximum is achievable when $\ket{\ps}=\ket{0\cdots 0}_{A_1\cdots A_n}\ox\ket{1\cdots 1}_{B_1\cdots B_n}$.
Eq. \eqref{eq:psps} implies that \eqref{eq:NcopyWernerSEPmin} holds for $n=1$. We prove  \eqref{eq:NcopyWernerSEPmin} by the induction. Suppose \eqref{eq:NcopyWernerSEPmin} holds for $n-1$, namely $\min_{\ps}\bra{\ps}(\r_s)^{\ox (n-1)}\ket{\ps}\ge{1 \over 24^{n-1}}$.
It is known that $\r_s$ is a two-qutrit separable Werner state of
rank eight \cite{dss00}. 
Let
$\r_s=\sum_i q_i\proj{a_i,b_i}$. We have
\bea
\label{eq:brapsrs}
&&
\bra{\ps}(\r_s)^{\ox n}\ket{\ps}
\notag\\
=&&
\bra{\ps}(\r_s)^{\ox (n-1)}\ox(\sum_i q_i\proj{a_i,b_i})_{A_nB_n}\ket{\ps}
\notag\\
=&&
\sum_i q_i
\norm{\braket{a_i,b_i}{\ps}}^2
\bra{\ps_i}(\r_s)^{\ox (n-1)}\ket{\ps_i}
\notag\\
\ge&&
\sum_i q_i
\norm{\braket{a_i,b_i}{\ps}}^2
{1 \over 24^{n-1}}
\notag\\
=&&
\bra{\ps}(\r_s)_{A_nB_n}\ket{\ps}
{1 \over 24^{n-1}}
\notag\\
\ge&&
{1 \over 24^{n}}.
\eea
Here $\ket{\ps_i}={\braket{a_i,b_i}{\ps} \over  \norm{\braket{a_i,b_i}{\ps}}}\in\cH_{A_1\cdots A_{n-1}}\ox\cH_{B_1\cdots B_{n-1}}$ is a normalized state of Schmidt rank at most two. The first inequality in \eqref{eq:brapsrs} follows from the induction assumption. To prove the last inequality in \eqref{eq:brapsrs} we assume that $\ket{\ps}=\sqrt{\l}\ket{w,x}+\sqrt{1-\l}\ket{y,z}$ is the Schmidt decomposition where
\bea
&&
\ket{w}=\sum_j \sqrt{\a_j} \ket{j,\e_j}_{A_1\cdots A_{n-1},A_n},
\\
&&
\ket{x}=\sum_k \sqrt{\b_k} \ket{k,\z_k}_{B_1\cdots B_{n-1},B_n},
\\
&&
\ket{y}=\sum_j \sqrt{\g_j} \ket{j,\i_j}_{A_1\cdots A_{n-1},A_n},
\\
&&
\ket{z}=\sum_k \sqrt{\d_k} \ket{k,\k_k}_{B_1\cdots B_{n-1},B_n},
\eea
and $\braket{w}{y}=\braket{x}{z}=0$. For $\ket{\ph_{jk}}:=\sqrt{\l\a_j\b_k} \ket{\e_j,\z_k} +  \sqrt{(1-\l)\g_j\d_k} \ket{\i_j,\k_k}$ and its normalization $\ket{\ph_{jk}'}$, we have
\bea
&&
\bra{\ps}(\r_s)_{A_nB_n}\ket{\ps}
\notag\\
&=&
\sum_{j,k} 
\bra{\ph_{jk}} \r_s \ket{\ph_{jk}}
\notag\\
&=&
\sum_{j,k} 
\norm{\ph_{jk}}^2
\bra{\ph'_{jk}} \r_s \ket{\ph'_{jk}}
\notag\\
&\ge&
\sum_{j,k} 
\norm{\ph_{jk}}^2
\min_{\m}
\bra{\m} \r_s \ket{\m}
\notag\\
&=&
\min_{\m}
\bra{\m} \r_s \ket{\m}
={1\over24},
\eea
where $\ket{\m}$ is an arbitrary bipartite pure state of Schmidt rank at most two. So the last inequality in \eqref{eq:brapsrs} holds.
This completes the proof.
 \epf

 \bl
 \label{le:undistillable}
For any integer $n$, and sufficiently small $\e=\e(n)>0$, the 
two-qutrit NPT state $\r$ in \eqref{eq:r=s} is $n$-undistillable. 
 \el
 \bpf
For any pure
state $\ket{\ps}$ of Schmidt rank two, we have
 \bea
\label{eq:braps}
 \bra{\ps}(\r^\G)^{\ox n}\ket{\ps}
 &=&
 \bra{\ps}(\s^\G-\e\proj{f^*,g})^{\ox n}\ket{\ps}
 \notag\\
 &:=&
 \bra{\ps} (\s^\G)^{\ox n} \ket{\ps} + \sum^n_{k=1} c_k \e^k,
 \eea
where $c_k$ are complex numbers. The definition of $\s$ implies 
that $\s^\G
\ge p_1 (I_9-\proj{\Ps}) \ge 0$ where $p_1$ is the minimal eigenvalue of
$\s^\G$. Hence
 \bea
 &&
\bra{\ps}(\r^\G)^{\ox n}\ket{\ps}
\notag\\
&\ge&
p_1^n \bra{\ps} (I_9-\proj{\Ps})^{\ox n} \ket{\ps} + \sum^n_{k=1} c_k \e^k.
 \eea
By Lemma \ref{le:NcopyWernerSEP}, the first summand is positive.
Since it has nothing to do with $\e$, the claim follows. 
 \epf

 \bcj
There is an undistillable two-qutrit NPT
state $\r$ in \eqref{eq:r=s} by a constant
$\e>0$.
 \ecj

\section{conclusion}
\label{sec:con}

We have proposed methods to detect the 1-distillability of two-qutrit NPT states under LOCC.
By applying them, we have proved that bipartite NPT states of rank four are 1-distillable. So they are useful resource for quantum-information tasks. We also have constructed bipartite NPT states of rank five which are not 1-distillable. We conjecture that they are not distillable, and so are related to the 
distillability problem. 
The next step is to investigate these states and study their 2-distillability.

\section*{Acknowledgments}

LC was supported by the NSF of China (Grant No. 11501024), and the Fundamental
Research Funds for the Central Universities (Grant Nos. 30426401 and 30458601). The second author was supported in part by an NSERC Discovery Grant.

\bibliographystyle{unsrt}


\end{document}